
\magnification=\magstep1
\baselineskip=0.822truecm
\centerline {\bf LINEAR POTENTIALS AND GALACTIC ROTATION CURVES}
\vskip 1.90truecm
\centerline {\bf Philip D. Mannheim}
\centerline {Department of Physics}
\centerline {University of Connecticut}
\centerline {Storrs, CT 06269-3046}
\smallskip
\centerline{mannheim@uconnvm.bitnet}
\vskip 3.00truecm
\centerline {\bf Abstract}
\vskip 0.60truecm
We present a simple, closed form expression for the potential of an
axisymmetric disk of stars interacting through gravitational potentials
of the form $V(r)=-\beta /r+\gamma r/2$, the potential associated with
fundamental sources in the conformal invariant fourth order theory of
gravity which has recently been advanced by Mannheim and Kazanas as a
candidate alternative to the standard second order Einstein theory. Using
the model we obtain a reasonable fit to some representative galactic
rotation curve data without the need for any non-luminous or dark matter.
Our study suggests that the observed flatness of rotation curves
might only be an intermediate phenomenon rather than an asymptotic one.
\vskip 3.00truecm
$~~~~~~~~~~~~~~$June, 1993$~~~~~~~~~~~~~~~~~~~~~~~~~~~~~~$UCONN-92-3
\vfill\eject
\hoffset=0.0truein
\hsize=6.5 truein
\noindent
{\bf (I) Introduction}
\medskip

In formulating a physical theory it is necessary to both work up from
phenomenological observations and down from fundamental principles, and to
be prepared to revise the insights obtained from both approaches as new
data come on line. However, after a program such as this has been
successfully carried through once, there is then some reluctance on the part
of the community to have to reopen the issue even in the light of subsequent
data.
Consequently, the prevailing view on galactic rotation curve data is that
their deviation from the behavior expected on the basis of the standard
Newton-Einstein theory as applied to the observed galactic luminous matter
surface brightness distribution must
be attributed to a (rather substantial) non-luminous or dark matter
galactic component. Since there is no clear evidence today that the
dominant component of the Universe is in fact non-luminous, there is thus
some merit in going back over familiar ground to see where, if anywhere,
something could be modified. (Milgrom (1983) and Sanders (1990) have also
looked
at possible revisions to the standard theory, with the recent study of
Begeman et al (1991) in particular showing that Milgrom's MOND alternative
is currently  phenomenologically viable).

Noting that there is currently no known theoretical reason which would select
out the standard second order Einstein theory from amongst the infinite class
of (all order)
covariant, metric based theories of gravity that one could in principle
at least consider, Mannheim and Kazanas have reopened the
question of what the correct covariant theory of gravity might be (Mannheim and
Kazanas (1989, 1991, 1992), Mannheim (1990, 1992, 1993), Kazanas and Mannheim
(1991)),
and
developed an approach which works down from an additional fundamental
principle,
namely that of local scale or conformal invariance, the invariance now believed
to be possessed by the other three fundamental strong, electromagnetic and weak
interactions. This invariance forces gravity to be described uniquely by the
fourth order action
$I_W = -\alpha \int d^4x (-g)^{1/2}
C_{\lambda\mu\nu\kappa}C^{\lambda\mu \nu\kappa}$
where $C_{\lambda\mu\nu\kappa}$ is the conformal Weyl tensor and $\alpha$
is a purely dimensionless coefficient.
In their original paper Mannheim
and Kazanas (1989) obtained the
exact, non-perturbative exterior vacuum solution associated with a static,
spherically symmetric gravitational source such as a star in this theory, viz.
$$-g_{00}= 1/g_{rr}=1-\beta(2-3 \beta \gamma )/r - 3 \beta \gamma
+ \gamma r - kr^2 \eqno(1)$$
where $\beta, \gamma,$ and $k$ are three appropriate dimensionful
integration constants. As they noted, this solution contains the familiar
exterior Schwarzschild solution and thereby yields the standard exterior
Newtonian potential term and its standard general relativistic corrections
whenever the two additional potential terms in Eq. (1) may be ignored.
The theory thus contains the same solution as the standard Einstein theory
in the appropriate kinematic regime even while not containing the Einstein
equations themselves, this being all that observation can require, and thus
nicely meets the demands of solar system scale observations.
The quadratic term in Eq. (1) may be associated with a general cosmological
background de Sitter geometry and is otherwise uneventful, and thus
(with both the  $3\beta\gamma$ terms being numerically
negligible - see below) the conformal theory
leads to the non-relativistic gravitational potential
$V(r)=-\beta/r+ \gamma r/2 $,
which may then be fitted to data whenever the weak gravity limit is
applicable.
$V(r)$ is thus the potential obtained in coming down
from a fundamental principles approach. In this paper we shall study the
implications of this potential by working up from galactic data, the first
distance
scale on which the potential is found to lead to significant deviations from
Newton.
As we shall see, the two approaches even have a chance to converge; however,
those
readers who may not be too comfortable with (or even disapprove of) the
whole general conformal gravity program can view this paper purely as an
attempt to identify which phenomenological potentials the currently
available observational data actually permit.

\noindent
{\bf (2) The Potential of an Extended Disk}

For an explicit application of conformal gravity to objects such as galaxies,
the
treatment completely parallels the Newton-Einstein treatment, i.e. the
potential of
each of the fundamental stellar constituents of the galaxy is first found by
solving
the relativistic theory exactly, with the potential of the galaxy then being
found
perturbatively by adding up the potentials of these stars in the weak gravity
limit.
Thus for our purposes here the requisite stellar potential is the above
$V(r)=-\beta/r+ \gamma r/2 $ with $\gamma$ being a new stellar parameter about
which
almost nothing is currently known ($\gamma$ is however known to in principle be
completely
independent of $\beta$ (Mannheim and Kazanas (1992))), with the
phenomenological
fitting to be presented below providing some first insight into typical
numerical values
for $\gamma$.

In order to actually determine the
weak gravity potential of an extended object such as a disk, we have
found it convenient to generalize the approach of Toomre (1963) first to
non-thin Newtonian disks (a step also taken by Casertano (1983)), and then to
disks with
linear potentials. The method which leans heavily on the completeness
properties
of the $J_n(kR)$ Bessel functions will be reported elsewhere, and here we state
only the relevant results. The Newtonian potential of a general
axisymmetric distribution of stellar matter sources
each with the same average $\beta$ (i.e. ignoring any morphological variation
throughout
the galaxy and treating each star as though it has the same typical $\beta$)
and with luminosity density function
$\rho(R,z^{\prime})$ is calculated to take the form
$$V_{\beta}(r,z)=-2\pi\beta\int_{0}^{\infty} dk\int_{0}^{\infty}dR
\int_{-\infty}^{\infty}dz^{\prime}
R \rho(R,z^{\prime})J_0(kr)J_0(kR) e^{-k\vert z -  z^{\prime}\vert}\eqno(2)$$
where $R,~z^{\prime}$ are cylindrical source coordinates
and $r$ and $z$ are the only observation point coordinates of relevance.
Analogously, the net contribution of a disk of stars each
with a linear $\gamma r/2$ potential (i.e. ignoring any possible morphological
variation in $\gamma$) is found to take the simple form
$$V_{\gamma}(r,z)=\pi\gamma\int dkdRdz^{\prime}R
\rho(R,z^{\prime})$$
$$\times~[(r^2+R^2+(z-z^{\prime})^2)J_0(kr)J_0(kR)
-2rR J_1(kr)J_1(kR)]
e^{-k\vert z -  z^{\prime}\vert}\eqno(3) $$
\noindent
For a thin exponential disk with $\rho(R,z^{\prime})=
\Sigma(R)\delta(z^{\prime})=\Sigma_0 \exp (-\alpha R)\delta(z^{\prime})$
where $R_0=1/\alpha$ is the disk scale length  and $N=2\pi\Sigma_0 R_0^2$
is the total number of stars in the disk, it is possible to perform the
integrations completely to yield
$$V_{\beta}(r)= -\pi\beta\Sigma_0 r[I_0(\alpha r/2)K_1(\alpha r/2)-
I_1(\alpha r/2)K_0(\alpha r/2)]\eqno(4)$$
for the Newtonian contribution (Freeman (1970)), and
$$V_{\gamma}(r)= \pi\gamma\Sigma_0 \{ (r/\alpha^2)[I_0(\alpha r/2)K_1(\alpha
r/2)-I_1(\alpha r/2)K_0(\alpha r/2)]$$
$$+ (r^2/2\alpha)[I_0(\alpha r/2)K_0(\alpha r/2)+
I_1(\alpha r/2)K_1(\alpha r/2)] \}
\eqno(5)$$
for the net linear contribution.
Differentiation then yields the remarkably compact expression
$$rV^{\prime}(r)=
(N\beta\alpha^3 r^2/2)[I_0(\alpha r/2)K_0(\alpha r/2)-
I_1(\alpha r/2)K_1(\alpha r/2)]$$
$$+(N\gamma\alpha r^2/2)I_1(\alpha r/2)K_1(\alpha r/2)
\eqno(6)$$
\noindent
an expression which behaves asymptotically as
$N\beta /r+N\gamma r/ 2 -3N\gamma R_0^2/ 4 r $
as would be expected. The coefficient
$N\beta$ is usually identified as $MG/c^2$ with $M$ being taken to be the
mass of the disk; and we see that for thin disks all departures from the
standard Freeman result are embodied in the $\gamma$-dependent term in the
simple and compact manner indicated.

\noindent
{\bf (3) Exponential Disks and Flat Rotation Curves}

In a recent comprehensive analysis of the $HI$ rotation curves of spiral
galaxies
(the more prominent $HII$ optical data studies of the type pioneered by Rubin
et. al.
(1978)
do not go out to a large enough number of scale lengths to show any
substantive deviation from a standard luminous Newtonian behaviour
(Kalnajs (1983), Kent (1986)) making the $HI$ data the main probe of the outer
reaches of the rotation curve),
Casertano and van Gorkom (1991) have found that the data fall into essentially
four general groups characterized by specific correlations between the maximum
rotation
velocity and luminosity; with the four groups being intermediate, compact
bright,
large bright, and dwarf galaxies. Thus as a first attempt at data fitting we
have chosen to study one representative galaxy from each group, specifically
the galaxies NGC3198, NGC2903, NGC5907, and DDO154. This will immediately
enable
us to test the flexibility of our theory, as well as confront the systematics
apparent in dark matter fits to the same four groups where it is typically
found
that the more luminous the galaxy the proportionately less dark matter
seems to be needed.

The intermediate galaxy NGC3198 is particularly well-suited for testing
theories since for it the data go out to
the largest known number of surface brightness scale lengths; and, with the
data being so flat, this galaxy is generally regarded as being prototypical.
To model the galaxy we have followed van Albada et. al. (1985), and
represented the surface brightness by a single exponential with a 1$^{\prime}$
(=2.72 kpc) scale length. (This choice approximates Wevers et. al. (1986)
$U^{\prime}$,
$J$, and $F$ filter data with eyeball slopes of $R_0=63 ^{\prime \prime},~
58 ^{\prime \prime},~54 ^{\prime \prime}$ respectively at a 5$\%$ uncertainty
level (the $F$ filter data have also been confirmed by Kent (1987)),
while ignoring a  spike in the very small angle  region data, and also a
possible truncation at the edge of the visible region). Moreover, the model
ignores
any modifications to the luminosity profile due to extinction or galactic dust
infrared reprocessing. Following Begeman (1987, 1989)
we have assigned a $z-$thickness to the disk according to the general analysis
of van der Kruit and Searle (1981), so that the luminosity density function
$\rho(R, z^{\prime})$ needed for Eqs. (2) and (3) takes the separable form
$\Sigma(R)$sech$^2(z^{\prime}/z_0)/2z_0$ with $z_0=R_0/5$.
(The $z-$thickness structure of the disk is only significant at small radii
where it serves to ensure that the inner part of the rotation curve is well
fitted by the Newtonian contribution, thus making it
possible to explore fully the effect of the linear potential on the outer
region).
Recognizing a 15$\%$ or so contribution to the
visible mass density from the $HI$ gas itself, we have also included the gas as
a matter source, and have found that, for model purposes, Wevers et. al. (1986)
$HI$ surface density data can be well represented by a sum of three
exponentials,
viz. $\sigma_{HI}(R)=(37.0 \exp (-R/2.23)+34.6 \exp (-R/0.87)-68.2 \exp
(-R/1.21)
)~M_{\odot}$/pc$^2$ ($R$ is in arc minutes) with a total $HI$ mass (to
infinity)
of $5.2\times10^{9}~M_{\odot}$,
of which $4.9\times10^{9}~M_{\odot}$ is observed in the explored
12$^{\prime}$
region. Finally, again following Begeman, and also van Albada and Sancisi
(1986),
we have multiplied the $HI$ gas profile by a
factor of 1.4 to allow for the presence of helium. With the model thus defined
with only the two free parameters $N$ and $\gamma$ of the stars,
we have generated the fit of Fig. (1) to Begeman's (1989) rotation curve
data. (The fitting proved not to be sensitive to any thickness for the gas so
we did
not use one, while the fit was also insensitive to any deviations of the value
of the
gaseous $\gamma/\beta$ ratio from the stellar one). With $N$
essentially being constrained by the overall normalization of the stellar
contribution, our best fit is found to yield a value of
$3.8\times10^{10}~M_{\odot}$ for the mass of the stars which is quite
reasonable for a galaxy with quoted luminosity
$L_B=9.0\times10^9L_{B\odot}$ (i.e. galactic mass to light ratio
$M/L=4.2~M_{\odot}/L_{B\odot}$),
with the obtained galactic mass in fact being
a typical so called maximum disk mass
in which the Newtonian term gets to be as large as it possibly can be.
Additionally, we find that the coefficient of
the net galactic linear term is given by $1/\gamma_{galaxy}=1/N\gamma_
{star}=2.9\times10^{29}$ cm to yield a galactic gamma to light ratio
$\gamma_{galaxy}/L_B=3.9 \times 10^{-40}$ /cm /$L_{B\odot}$.
As we can see from Fig. (1),
the contribution of the linear potential piece is
remarkably reminiscent in shape to that of a typical dark matter contribution
to
galactic data fitting (see e.g. Kent (1987) for an extensive study); and
intriguingly we find that the linear potential is  competitive with the
Newtonian one in a galaxy when $1/\gamma_{galaxy}$
is of order the Hubble radius, the naive anticipation of
Mannheim and Kazanas (1989) in their original study. Given this value for
$\gamma_{galaxy}$, the inferred
value for $\gamma_{star}$ is then $0.9 \times 10^{-40}$ cm$^{-1}$, making the
linear
potential indeed negligible on solar system distance scales as initially
required,
with the linear potential only first becoming competitive with the Newtonian
one galactically. (In passing we note that
with such a small value
for $\gamma$ the $\beta\gamma$ product terms in Eq. (1) are then rendered
completely insignificant, a fact we had indicated earlier).
As regards an assessment of the quality of our fitting, we should note
that there are
some still not fully understood discrepancies
in the outer region (of order up to
7 km/s) between Begeman's data and Bosma's earlier 1978, 1981 data;
while additionally Begeman quotes a maximum difference
of 3 km/s between his inner rotation curve
and that of Hunter et al (1986). Moreover,
we should also note that Begeman's last 2 data points (the farthest) actually
use adopted
values which are extrapolated from closer in ones. Also, there is even some
indication
in the data of a warp at the largest observed distances which we have not
attempted to model. Finally, in general as regards rotation curve
data, we note that velocities are measured at corresponding
distances on the two sides
(receding and approaching) of the galaxy, with differences between
the obtained values usually constituting the (sole) quoted errors,
even though
such differences could well indicate that the galaxy actually has a
symmetry lower than that of a disk (in which case the very extraction of the
original mass density itself from the surface brightness data becomes
slightly suspect).
Thus fitting at the 5$\%$ (or even 10$\%$) level would seem to be
acceptable, making our fitting quite adequate.

For the compact, bright galaxy
NGC2903 Wevers et al (1986) find a stellar disk scale length of 2.0 kpc
(=1.08$^{\prime}$). For this galaxy the contribution of the gas is quite small
($M_{HI}=2\times10^{9}~M_{\odot}$), and introducing a stellar
or a gaseous thickness was found to have no appreciable effect.
Consequently we are able to fit the galaxy directly with Eq. (6)
(using the same $\gamma/\beta$ ratio for stars and gas) to find the fit of
Fig. (1) to Begeman's (1987) rotation curve data data. We find a fitted mass of
$5.3\times10^{10}~M_{\odot}$ for the stars to be compared with quoted
luminosity
$L_B=1.5\times10^{10}L_{B\odot}$ ($M/L=3.5~M_{\odot}/L_{B\odot}$), and a
net galactic linear term given by $1/\gamma_{galaxy}=1/N\gamma_
{star}=1.3\times10^{29}$ cm, making $\gamma_{star}=1.4 \times 10^{-40}$
cm$^{-1}$
and $\gamma_{galaxy}/L_B=5.1 \times 10^{-40}$ /cm /$L_{B\odot}$.
As regards the fitting we note that this galaxy also has a warp at large
distances
which we have not attempted to model. Also, in his original dark matter fit
Begeman
(1987) found a lot of scatter in the inner rotation curve (he actually settled
for
an eye-ball fit rather than a least squares one), prompting him to suggest that
there
might be an additional stellar component in the not well explored close in
region.
If there is, then in the present
theory such a component would also contribute in the outer region because of
the
linear term.

Until very recently the large, bright galaxy NGC5907 had actually had a severe
fitting
problem because
the original surface brightness data of van der Kruit and Searle (1981) were
simply
completely incompatible with the rotation curve data
(van Albada and Sancisi (1986)) in the inner region where the stellar Newtonian
potential should dominate. This
situation has only recently been rectified by Barnaby and Thronson (1992a) who
find
(using a different filter) a completely different surface brightness curve, one
which
does nicely fit
the inner region (Barnaby and Thronson (1992b)). Barnaby and Thronson find that
the stellar disk is well parametrized by
$\rho(R, z^{\prime})=\Sigma_0 \exp (-R/R_0)$sech$(z^{\prime}/z_0)/\pi z_0$
where $R_0$ =4.0
kpc (=1.22$^{\prime}$), and $R_0/z_0=9.2$. Additionally they find a close in
(and thus
easy to miss) central region stellar bulge with a luminosity
5$\%$ of that of the disk which they parametrize by a modified Hubble profile
($\sim 1/(R^2+R_0^2)$) with scale length $R_0$ =0.07$^{\prime}$. The
total stellar contribution to the rotation curve
is exhibited in Fig. (1). We find the mass of the disk to be
$1.1\times10^{11}~M_{\odot}$
and that of the bulge to be $1.7\times10^{10}~M_{\odot}$ with the quoted
luminosity
of the galaxy being $L_B=1.8\times10^{10}L_{B\odot}$
($M/L=6.1~M_{\odot}/L_{B\odot}$ for the disk).
For the linear term we find
(we use the same $\gamma/\beta$ ratio for the disk and bulge)
$1/\gamma_{galaxy}=1/N\gamma_
{star}=1.7\times10^{29}$ cm, making $\gamma_{star}=5.5 \times 10^{-41}$
cm$^{-1}$
and $\gamma_{galaxy}/L_B=3.2 \times 10^{-40}$ /cm /$L_{B\odot}$,
to give values which are comparable with those of the other galaxies.

For the dwarf irregular DDO154 Carignan and Freeman (1988) and Carignan and
Beaulieu
(1989) have determined both the rotation curve and the surface brightness data.
The stellar component is fit by a disk with scale length 0.43$^{\prime}$
(corresponding
to
$R_0=0.5$ kpc if the galaxy is at a distance $D$=4 Mpc - it may be at $D$=10
Mpc,
see below); while the gas is well fitted by
$\sigma_{HI}(R)=(31.6 \exp (-R/1.42)-25.7 \exp (-R/1.08))~M_{\odot}$/pc$^2$
($R$ is in arc minutes) with a total $HI$ mass (to infinity)
of $2.8\times10^{8}~M_{\odot}$ (at $D$=4 Mpc), 93$\%$  of which is observed in
the
explored region. Since the gas turns out to be the main gravitational component
we see that the observed region corresponds to about 4.5
leading 1.4$^{\prime}$ gas scale lengths. In the
central galactic region the stellar surface brightness is
not yet fully explored (it actually
appears to be flattening off, meaning that an exponential disk could be
overestimating
the inner surface brightness). Consequently in order to fit the inner rotation
curve we
have
found it necessary to give the stellar component a
sech$^2(z^{\prime}/z_0)/2z_0$
thickness with $z_0=R_0$(stellar)/3.
Additionally, we have allowed the stellar and gaseous $\gamma/\beta$ ratios
to vary independently (there is no immediate reason why
they should be the same,
anymore than the mass to light ratios of bulges and disks of a given
galaxy should be the same). We have not included the contribution of gas
pressure
(a 5$\%$ or so effect which is completely ignorable in galaxies where stars
dominate the dynamics), nor have we considered any random motions of the gas
(a 1 or 2$\%$ effect), or an apparent galactic warp. Also we note that
the very last rotation curve data point was only determined on one
side (the receding one) of the
galaxy and its inferred velocity is not independent of those of the two
immediately
previous points.
Our best fit at $D$=4 Mpc is presented in Fig. (1) with stellar
mass $6.8\times10^{7}~M_{\odot}$ for a quoted luminosity at that distance of
$L_B=5.0\times10^7L_{B\odot}$ ($M/L=1.4~M_{\odot}/L_{B\odot}$).
For the stars we find
$1/\gamma_{galaxy}=1/N\gamma_
{star}=4.0\times10^{29}$ cm, making $\gamma_{star}=3.7\times 10^{-38}$
cm$^{-1}$
and $\gamma_{galaxy}/L_B=5 \times 10^{-38}$ /cm /$L_{B\odot}$,
while the best value for $\gamma_{gas}$ is found to be zero (i.e. much smaller
than
the stellar contribution). Now while Carignan and coworkers favor putting
DDO154 at
$D$=4 Mpc, we note that they also indicated
a possible adopted distance at $D$=10 Mpc which Krumm and Burstein (1984)
favor, this
being the distance at which the Tully-Fisher relation is obeyed. Since the
amount of
gas is 6.25 times bigger at the larger distance, it would then be totally
dominant.
(While we shall continue to use the same
$\sigma_{HI}(R)$ as before after scaling up to the larger distance,
we note that Krumm and Burstein actually obtained
a leading scale length of $2.5^{\prime}$). At $D$=10 Mpc the fit is found to be
insensitive to any stellar or gaseous disk thickness, with the thin disk
approximation then yielding the fit of Fig. (1).
We find the stellar mass to be $7.4\times10^{7}~M_{\odot}$.
Additionally we find that for the stars
$1/\gamma_{galaxy}=1/N\gamma_
{star}=1.4\times10^{30}$ cm, making $\gamma_{star}=9.9 \times 10^{-39}$
cm$^{-1}$,
while the best value for $\gamma_{gas}$ is again found to be zero. (Actually,
the fits at both the candidate distances can even be improved in the outer
region
if we allow $\gamma_{gas}$ to go negative. This would be somewhat difficult
to understand, though without a dynamical theory for $\gamma$ it cannot yet
be excluded). Since our theory, the standard flat dark matter theory, and MOND
would all eventually overshoot the data if the suggested
large distance decline in the DDO154 rotation
curve were to be
confirmed, further observational study of this point might
prove interesting.

In our fitting we see a reflection of the general luminosity trend
found in dark matter fits, with both the relative Newtonian contribution
and the inferred galactic mass to light ratios increasing
with luminosity. Additionally, and intriguingly we find that the
values obtained for $1/\gamma_{galaxy}$ are remarkably close to each other and
to
the Hubble radius suggesting some possible
common underlying dynamics. (Perhaps $\gamma_{galaxy}$ sets or is set by the
scale at
which galaxies can fluctuate out of an initial cosmological background).
For the three regular galaxies NGC3198, NGC2903, and NGC5907 we find that the
mass to
light ratios are comparable to each other and likewise their gamma to light
ratios, to
thus suggest only a mild morphological dependence to the average stellar
$\beta$ and
$\gamma$ parameters used as input for Eqs. (2) and (3). (While the mass to
light
ratio is assumed to be uniform within a given galaxy, the actual numerical
value of this
ratio is not apparently universal for all galaxies suggesting that galaxies do
not all
have the same universal mix of stars and/or the same typical average stellar
$\beta$.
A similar situation should thus be expected
to obtain for the galactic gamma to light ratio). We note that the irregular
galaxy DDO154
does not show the same galactic gamma to light ratio as the other three
galaxies. It is
not clear
whether this is a fundamental issue or whether perhaps the galaxy has an
anomalous mix of
stars. As regards this issue  we recall that both the dark matter and MOND fits
of
Begeman et al (1991) to the same galaxy find atypically small mass to
light ratios. Thus dwarf irregulars may be fundamentally different, though of
course
trying to extract out stellar parameters in a galaxy where the stars do not
dominate
may not be completely reliable.
With regard to our fitting, we see that while our fitting is
yielding flat rotation curves, it is doing so in a theory in which rotation
curves must
eventually rise. This stands in marked contrast to the
asymptotically flat behaviour expected both in MOND
and in the isothermal gas sphere model of dark matter with its asymptotically
logarithmic
galactic potential. (Other, less popular, asymptotic alternatives for dark
matter
are considered in van Albada et al (1985)). Since one may unfortunately
run out of
galaxy before possibly seeing any such rise, perhaps
the sharpest difference between linearly rising and logarithmic potentials may
emerge at
the slightly larger distance scale associated with
gravitational lensing where such differences might
even be pronounced, thus making a study of
the (so far unknown) conformal gravity lensing
predictions in the non-asymptotically flat geometry of Eq. (1) quite urgent.

We believe that we have
thus established the candidacy (at least)
of fourth order gravity by working up from the
non-relativistic limit; and since the conformal theory has already been
shown to possess no flatness problem (Mannheim (1992)) and thus not require
any cosmological dark matter, and since the linear potential has also been
shown
to be capable of yielding
galactic stability without the need for dark matter (Christodoulou (1991)),
we see that both cosmologically and
galactically it might turn out to be the case that luminous matter is the
major constituent of the Universe after all.

The author would like to thank D. Kazanas, D. Christodoulou and J. Taylor
for stimulating discussions, and D. Barnaby, K. Begeman, C. Carignan, S.
Casertano,
S. Kent and R. Sancisi for some very helpful communications.
This work has been supported in part
by the Department of Energy under grant No. DE-FG02-92ER40716.00.
\vfill\eject

\noindent
{\bf References}

\noindent Barnaby, D., and Thronson, H. A. 1992a, A. J. 103, 41.

\noindent Barnaby, D., and Thronson, H. A. 1992b, B.A.A.S. 24, 809.

\noindent Begeman, K. G. 1987, Ph. D. Thesis, Gronigen University.

\noindent Begeman, K. G. 1989, A. A., 223, 47.

\noindent Begeman, K. G., Broeils, A. H., and Sanders, R. H. (1991),
Mon. Not. R. Astron. Soc. 249, 523.

\noindent Bosma, A. 1978, Ph. D. Thesis, Gronigen University.

\noindent Bosma, A. 1981, A. J., 86, 1791.

\noindent Carignan, C., and Freeman, K. C. 1988, Ap. J. (Letters) 332, 33.

\noindent Carignan, C., and Beaulieu, S. 1989, Ap. J. 347, 760.

\noindent Casertano, S. 1983, Mon. Not. R. Astron. Soc. 203, 735.

\noindent Casertano, S., and van Gorkom, J. H. 1991, A. J., 101, 1231.

\noindent Christodoulou, D. M. 1991, Ap. J., 372, 471.

\noindent Freeman, K. C. 1970, Ap. J. 160, 811.

\noindent Hunter, D. A., Rubin, V. C., and Gallagher III, J. S. 1986, A. J. 91,
1086.

\noindent Kalnajs, A. J. 1983, in Internal Kinematics and Dynamics of Disk
Galaxies,
IAU Symposium No. 100, ed. E. Athanassoula (Reidel, Dordrecht), p. 87.

\noindent Kazanas, D., and Mannheim, P. D. 1991, Ap. J. Suppl.
Ser., 76, 431.

\noindent Kent, S. M. 1986, A. J. 91, 1301.

\noindent Kent, S. M. 1987, A. J. 93, 816.

\noindent Krumm, N., and Burstein, D. 1984, A. J. 89, 1319.

\noindent Mannheim, P. D. 1990, Gen. Rel. Grav., 22, 289.

\noindent Mannheim, P. D. 1992, Ap. J., 391, 429.

\noindent Mannheim, P. D. 1993, Gen. Rel. Grav. (in press).

\noindent Mannheim, P. D., and Kazanas, D. 1989, Ap. J., 342, 635.

\noindent Mannheim, P. D., and Kazanas, D. 1991, Phys. Rev., D44, 417.

\noindent Mannheim, P. D., and Kazanas, D. 1992, Newtonian limit of
conformal gravity and the lack of necessity of the second order Poisson
equation, UCONN-92-4, unpublished.

\noindent Milgrom, M. 1983, Ap. J. 270, 365; 371, 384.


\noindent Rubin, V. C., Ford W. K., and Thonnard, N. 1978, Ap. J. (Letters),
225, L107.

\noindent Sanders, R. H. 1990, A. A. Rev., 2, 1.

\noindent Toomre, A. 1963, Ap. J., 138, 385.

\noindent van Albada, T. S., Bahcall, J. N., Begeman, K. G., and Sancisi,
R., 1985, Ap. J., 295, 305.

\noindent van Albada, T. S., and Sancisi, R.,
1986, Phil. Trans. R. Soc., A320, 447.

\noindent  van der Kruit, P. C., and Searle L., 1981,
A. A., 95, 105.

\noindent Wevers, B. M. H. R., van der Kruit, P. C., and Allen, R. J. 1986,
A. A. Suppl. Ser., 66, 505.
\medskip
\noindent
{\bf Figure Caption}

\noindent
Figure (1). The calculated rotational velocity curves
associated with the metric of Eq. (1) for the four representative galaxies,
the intermediate sized NGC3198, the compact bright NGC2903,
the large bright NGC5907, and the dwarf irregular DDO154 (at two possible
adopted distances). In each graph
the bars show the data points with their quoted errors, the full
curve shows the overall theoretical velocity prediction (in km/s)
as a function of distance (in arc minutes) from the center of each galaxy,
while the two indicated dotted curves show the rotation curves that the
separate Newtonian and linear potentials of Eq. (1) would produce when
integrated
over the luminous matter distribution of each galaxy. No dark matter is
assumed.
\end